\documentclass[12pt, letterpage]{article}
\usepackage{amssymb,latexsym}
\usepackage{natbib}
\usepackage[unicode=true, bookmarks=true, bookmarksnumbered=false, bookmarksopen=false, breaklinks=true, pdfborder={0 0 0},backref=false]{hyperref}
\hypersetup{
  colorlinks=true,
  allcolors=blue
}
\usepackage{amsmath}
\usepackage{bbold}
\usepackage{graphicx}
\usepackage{cases}
\usepackage{float}
\usepackage{subcaption}
\usepackage{multirow}
\usepackage{xcolor}
\usepackage{indentfirst,geometry,setspace,fancyhdr,sectsty}

\allowdisplaybreaks

\newtheorem{lemma}{Lemma}
\newtheorem{proposition}{Proposition}
\newtheorem{definition}{Definition}

\def\R{\mathbb{R}}

\begin{document}
\title{\LARGE  Institutional Screening and the Sustainability of Conditional Cooperation}
\vspace{1 cm}
\author{Ethan Holdahl\footnote{Department of Economics, University of California, Irvine. Email: eholdahl@uci.edu.} and Jiabin Wu\footnote{Department of Economics, University of Oregon. Email: jwu5@uoregon.edu.}
}

\date{\vspace{0.5 cm} \today \\
\vspace{-0.5 cm}
}
\maketitle

\begin{abstract}
This paper studies a preference evolution model in which a population of agents are matched to play a sequential prisoner's dilemma in an incomplete information environment. An institution can design an incentive-compatible screening scheme, such as a special zone that requires an entry fee, or a costly label for purchase, to segregate the conditional cooperators from the non-cooperators. We show that institutional intervention of this sort can help the conditional cooperators to prevail when the psychological benefit of cooperating for them is sufficiently strong and the membership of the special zone or the label is inheritable with a sufficiently high probability. 

\vspace{0.2cm}

\end{abstract}

\renewcommand*{\thefootnote}{\arabic{footnote}}
\setcounter{footnote}{0}

%\begin{f_0}
%\fontsize{12}{18}\selectfont
%\end{f_0}

%\baselineskip17pt

%\begin{spacing}{1.5}

\newpage

\section{Introduction}
Conditional cooperation has interested researchers across various disciplines for decades \citep{Ostrom2000JEP}. Human-subject experiments \citep[and many others]{FGF2001EL, FG2010AER} have provided ample evidence for the existence of conditional cooperation. It refers to the phenomenon that people cooperate in dilemma situations only when they believe that others also cooperate. Conditional cooperation can be explained by people's other regarding preferences such as reciprocal preferences \citep{rabin1993aer, DufwenbergKirchsteiger2004GEB}. 

Initiated by \cite{guthyaari1992} and \cite{guth1995ijgt}, the indirect evolutionary approach has become a workhorse model for studying the evolution of other-regarding social preferences \citep[see][for a survey]{AlgerWeibull2019ARE}. It assumes that evolution takes place at the preference level. Preferences determine behavior, which affects people's material payoffs. Material payoffs in turn determine the evolution of preferences. The indirect evolutionary approach thus offers a way to understand the evolutionary roots of various behavioral level phenomena, including conditional cooperation, from the preference level.
%The general lesson from the literature on indirect evolutionary approach is that non-materialistic preferences can survive under complete information but not under incomplete information when people in the population is randomly matched. 
In particular, \cite{guthkliemt1998RS} propose a model on the evolution of trustworthiness and show that its survival must be accompanied by conditional cooperation. When preferences are observable, the trustworthy type prevail in the long run because the trustworthy agents only cooperate when their opponents are trustworthy. In this case, the trustworthy agents act as conditional cooperators. However, when preferences are not observable, the share of trustworthy people is bound to decline because they can no longer behave differently accordingly to their opponents' types.

It is desirable for the conditional cooperators to differentiate themselves from the non-cooperators. This not only enables the conditional cooperators to behave differently when matching with different opponents, but also allows them to match only with other conditional cooperators. How can they achieve this when preferences are not observable? In this paper, we consider a potential solution: institutional screening. Institutional screening has been explored in the context of religion. For example, \cite{Iannaccone1992JPE} and \cite{CarvalhoSacks2021JPUBE}, among others, theoretically show how religious groups can use specific traits and rituals to screen out non-believers. The idea that institution can foster cooperation in dilemma situations can be traced back to Thomas Hobbes' Leviathan. We assume that a society consists of a population of people, including conditional cooperators and non-cooperators who are driven by different preferences, and an institution. Following \cite{BisinVerdier2021WP}, we conceptualize the institution as a mechanism through which social choices are delineated and implemented. This view on institutions is in line with \citep{North1981, North1990, David1994, Grief2006}. If differentiating the conditional cooperators from the non-cooperators serves the institutional goal, the institution may want to implement some screening mechanism to achieve it. For example, the institution can set up a special zone that requires entry fee or create an artificial label for people to purchase. The entry fee to the special zone or the cost of label should be designed in an incentive compatible way so that only the conditional cooperators are willing to pay for it. We show that such an entry fee/cost exists, which helps segregating the conditional cooperators from the non-cooperators. We show that whether the conditional cooperators can prevail in the long run depends on two critical features. First, the psychological benefit of cooperating for the conditional cooperators is sufficiently high, which guarantees that the institution always has an incentive to implement a screening scheme. Second, the membership of the special zone or the label is inheritable to a sufficiently high degree. That is to say that descendants of those with a membership face a lower membership cost than those whose parents were not members themselves. For example, assume the special zone is in a geographically remote location. Descendants of the special zone will be born there and wouldn't have to pay transportation costs to get to the special zone, thus reducing the effective membership cost. This discounted cost lowers the burden of future generations of conditional cooperators which is necessary to ensure their relative material payoffs remain competitive with the rest of the population.

This paper considers endogenous matching as opposed to the majority of the indirect evolutionary approach literature which considers exogenously random matching. A few papers consider matching being determined by institutions or democratic processes in the complete information environment \citep{RigosNax2016JTB, Wu2016EB, Wu2022WP}. We instead consider the role of institutions in determining matching in the incomplete information environment. Since the institutions cannot directly manipulate the matching patterns in the incomplete information environment, we utilize the concept of stable matching to describe how agents are matched after the screening stage. \cite{Wu2021RED}, \cite{Hilleratal2022WP} and \cite{WangWu2022WP} are a few recent attempts to introduce stable matching to evolutionary models. 

The paper is organized as follows. Section \ref{sec:model} lays out the setting of the model. Section \ref{sec:analysis} first re-illustrates the classic results without institutional intervention, then introduces institutional screening and studies its implications on the evolution of conditional cooperation. Section \ref{sec:conclusion} concludes. All proofs are relegated to the Appendix. 
\section{The Model} \label{sec:model}
\subsection{Basic Setup} \label{sec:setup}
\begin{figure}[!ht]%
\caption{A sequential prisoner's dilemma game with material payoffs \label{figure: prisoner's dilemma}}
\centering
\includegraphics[scale=0.4]{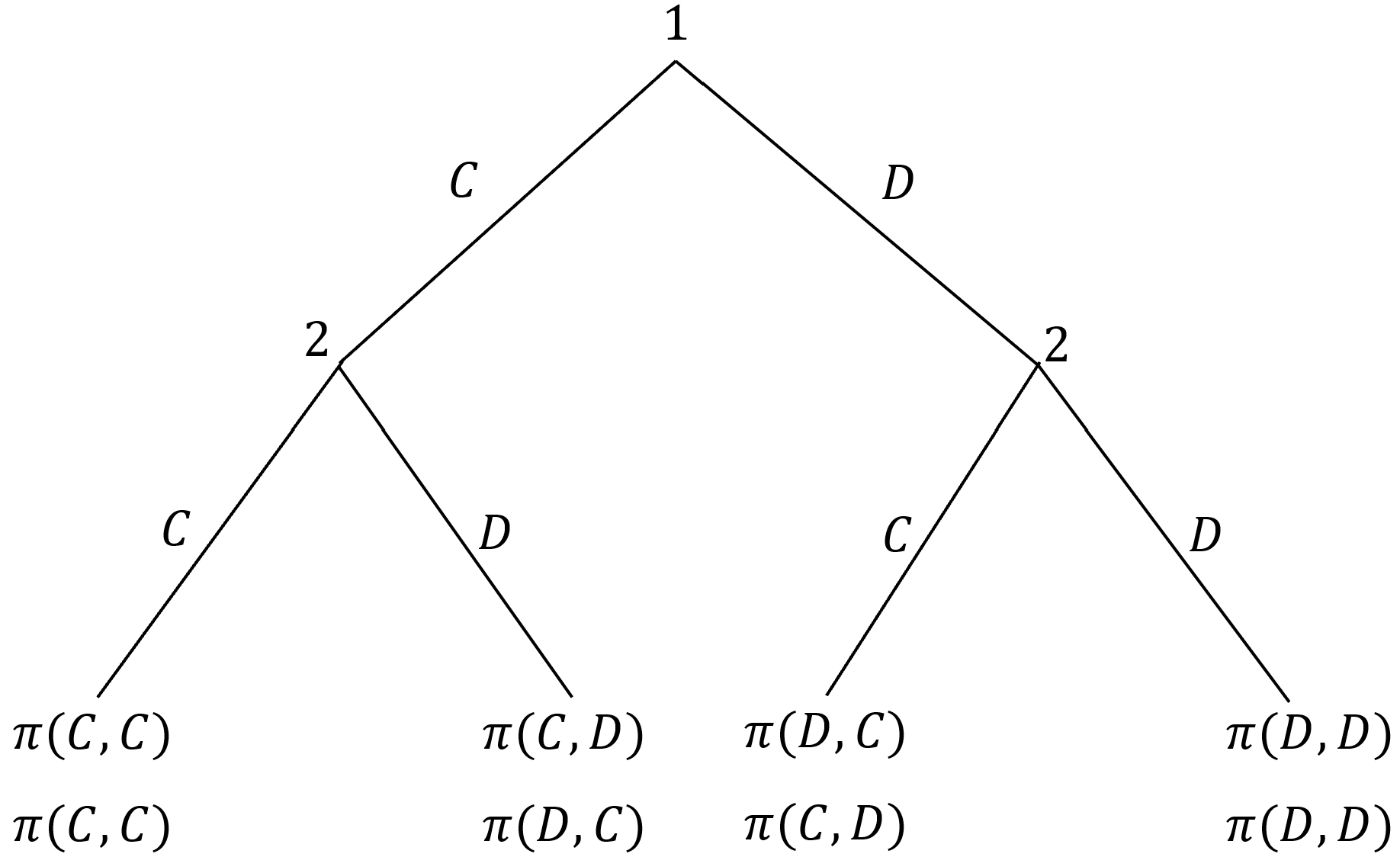}
\end{figure}

Consider a continuous population of agents of two preference types: $\theta$ and $\tau$. The amount of $\theta$-type agents is $x> 0$ and the amount of $\tau$-type agent is $y > 0$. $(x, y)$ denote the population state.  The agents are randomly matched in pairs to play a one-shot sequential prisoner's dilemma game as shown in Figure \ref{figure: prisoner's dilemma}. In each pair, each agent takes the role of player 1 with probability 0.5 independent of their types.

In the game, each agent has two actions: $\{C, D\}$. The game generates material payoffs for the agents. The material payoff function is given by $\pi(\cdot, \cdot): \{C, D\}^2 \rightarrow \R$, which satisfies $\pi(D, C)>\pi(C, C)>\pi(D, D)>\pi(C, D)$, the standard criterion for a prisoner's dilemma. The $\tau$-type agents' have materialistic preferences, that is, their preferences can be represented by the material payoff function. The $\tau$-type agents always choose $D$ as player 2, as it is their strictly dominant strategy. Hence, the $\tau$-type agent can be regarded as non-cooperators. A $\theta$-type agent's preferences are represented by the utility function $U_\theta(\cdot, \cdot): \{C, D\}^2 \rightarrow \R$. The utility function satisfies $U_\theta(C, C)=\pi(C, C)+\alpha$, where $\alpha>\pi(D, C)-\pi(C, C)$; and $U_\theta(s, s')=\pi(s, s')$, for $(s, s') \ne (C, C)$. In words, a $\theta$-type agent gains an extra psychological benefit of $\alpha$ only when choosing $C$ against another agent choosing $C$. Given the condition imposed on $\alpha$, a $\theta$-type agent as player 2 would choose $C$ upon observing that player 1 chooses $C$, and she would chooses $D$ upon observing that player 1 chooses $D$. Hence, the $\theta$-type agents can be regarded as conditional cooperators.

\subsection{Evolution} \label{evolution}
We first add time index to the population state: $(x_t, y_t)$ denotes the population state at time $t \in \mathbb{N}$ (assume that $x_0>0$ and $y_0>0$). Let $F_\theta(x_t, y_t)$ and $F_\tau(x_t, y_t)$ denote the average material payoffs of the $\theta$-type agents and the $\tau$-type agents at time $t$, respectively. We also define $\bar{U}_\theta(x_t, y_t)$ and $\bar{U}_\tau(x_t, y_t)$ as the average utility of the $\theta$-type agents and the $\tau$-type agents at time $t$, respectively. 

The population evolves across generations over time. The average material payoff of a type determines the reproductive rate of the agents of that type. Importantly, the evolution is governed by the comparison of material payoffs instead of utilities. When the average material payoff of the $\theta$-type agents is higher than that of the $\tau$-type agents in the current generation, the $\theta$ group grows faster than the $\tau$ group. The following model describes the evolutionary process:
\begin{eqnarray}
&&x_{t+1}=x_tF_\theta(x_t, y_t), \\
&&y_{t+1}=y_tF_\tau(x_t, y_t).
\end{eqnarray}

\begin{definition}\label{def: steady states}
We say $\theta$-type (respectively, $\tau$-type) \textbf{\textit{dominates}} the population  as $t$ approaches infinity. when we have $\lim_{t \rightarrow \infty}x_t/y_t=\infty$ (respectively, $\lim_{t \rightarrow \infty}y_t/x_t=\infty$.) We say the population reaches a \textbf{\textit{steady state}} at time $t$ if all agents in the population have the same reproduction rate from $t$ on, this requires $F_\theta(x_{t+k}, y_{t+k})=F_\tau(x_{t+k}, y_{t+k})=c$, for some constant $c>0$ and any integer $k\ge 0$. 
\end{definition}
\section{Analysis} \label{sec:analysis}
\subsection{Complete Information} \label{sec:complete}
In this section, we consider the benchmark case with complete information. That is, the agents' types are observable. In line with the literature on the indirect evolutionary approach, we first assume that agents in the population are uniformly randomly matched. Table \ref{table:SPE} summarizes all the subgame perfect equilibria (SPEs). One can observe that the equilibrium predictions depend solely on player 2's type. When player 2 is a conditional cooperator ($\theta$-type), both players would choose $C$ in equilibrium; when player 2 is a non-cooperator ($\tau$-type), both players would choose $D$ in equilibrium. 
\begin{table}
\caption{Subgame perfect equilibria under complete information} \label{table:SPE}
\begin{center}
\begin{tabular}{|c| c| c|} 
 \hline
 Player 1 & Player 2 & SPE \\
 \hline
 $\theta$ & $\theta$ & $(C, (C, D))$ \\ 
 \hline
 $\theta$ & $\tau$ & $(D, (D, D))$ \\ 
 \hline
 $\tau$ & $\theta$ & $(C, (C, D))$ \\ 
 \hline
$\tau$ & $\tau$ & $(D, (D, D))$ \\ 
 \hline
\end{tabular}
\end{center}
\end{table}

The average material payoffs of the two types are given by:

\begin{eqnarray}
F_\theta(x_t, y_t)&=&\frac{x_t}{x_t+y_t}\pi(C, C)+\frac{y_t}{x_t+y_t}(\frac{1}{2}\pi(D, D)+\frac{1}{2}\pi(C, C)), \\
 F_\tau(x_t)&=&\frac{x_t}{x_t+y_t}(\frac{1}{2}\pi(C, C)+\frac{1}{2}\pi(D, D))+\frac{y_t}{x_t+y_t}\pi(D, D).
\end{eqnarray}

%%%%%
%\begin{eqnarray}
%F_\theta(x_t, y_t)&=&\frac{x_t}{x_t+y_t}\pi(C, C)+\frac{x_t}{x_t+y_t}(\frac{1}{2}\pi(D, D)+\frac{1}{2}\pi(C, C)), \\
% F_\tau(x_t)&=&\frac{x_t}{x_t+y_t}(\frac{1}{2}\pi(C, C)+\frac{1}{2}\pi(D, D))+\frac{x_t}{x_t+y_t}\pi(D, D).
%\end{eqnarray}
%%%%%

Since $\pi(C, C)>\pi(D, D)$, we have $F_\theta(x_t, y_t)>F_\tau(x_t, y_t)$ for $x_t>0$, implying that $\theta$-type dominates the population as $t$ approaches infinity for any initial condition $(x_0, y_0)$. The rationale is that a conditional cooperator always cooperates with another conditional cooperator regardless of her role, but only cooperates with a non-cooperator when she is player 2. This makes sure that the conditional cooperators grab the benefit of cooperation among themselves and at the same time protect themselves from the non-cooperators' exploitation.

Since the types are observable, it is reasonable to assume that the agents can match voluntarily with those they prefer to match with. To capture this idea, we introduce the notion of stable matching. 

A matching configuration is a vector $\mu(x_t, y_t)=(\mu_{a, b}(x_t, y_t))$, where $\mu_{a, b}(x_t, y_t)$ denotes the amount of type-$a$ agents that are matched with type-$b$ agents, with $a, b \in \{\theta, \tau\}$ that satisfies the following requirements. 
\begin{eqnarray}
&& \mu_{\theta, \theta}(x_t, y_t)+\mu_{\theta, \tau}(x_t, y_t)=x_t, \ \mu_{\tau, \theta}(x_t, y_t)+\mu_{\tau, \tau}(x_t, y_t)=y_t, \nonumber \\
&& \text{and } \mu_{\theta, \tau}(x_t, y_t)=\mu_{\tau, \theta}(x_t, y_t).
\end{eqnarray}

We assume that the agents will reach a stable matching within each time period before they play the prisoner's dilemma. Since we are looking at a continuous population with finite types, we adopt the definition of stable matching for aggregate matching from \cite{Echeniqueetal2013ECMA} to our context, which serves as a natural generalization of the stability concept by \cite{GaleShapley1962AMM}. Let $K(a, b): \{\theta, \tau\}^2 \rightarrow R$ denote the expected equilibrium utility of a type-$a$ agent against a type-$b$ agent. 

\begin{definition}\label{def: stable matching}
A pair of types $(a, b)$, with $a, b \in\{\theta, \tau\}$, is a \textbf{blocking pair} for $\mu(x_t, y_t)$ if for $c \neq a$, $d \neq b$ and $c, d \in \{\theta, \tau\}$, we have $K(a, b)>K(a, d)$, $K(b, a)>K(b, c)$,  $\mu_{a, d}(x_t, y_t)>0$ and $\mu_{b, c}(x_t, y_t)>0$. A matching configuration $\mu(x_t, y_t)$ is \textbf{stable} if there are no blocking pairs for $\mu(x_t, y_t)$. 
\end{definition}

Lemma \ref{lemma:PAM} shows that perfectly assortative matching (all $\theta$-type agents are matched with one another, and all $\tau$-type agents are matched with one another) is the unique stable matching:

\begin{lemma} \label{lemma:PAM}
The unique stable matching is $\mu_{\theta, \theta}(x_t, y_t)=x_t, \mu_{\tau, \tau}(x_t, y_t)=y_t$.
\end{lemma}

Given that the agents are perfectly segregated through matching, the average material payoffs of the two types are given by $F_\theta(x_t)=\pi(C, C)$ and $F_\tau(x_t)=\pi(D, D)$. Since $\pi(C, C)>\pi(D, D)$, we have $F_\theta(x_t, y_t)>F_\tau(x_t, y_t)$ for $x_t>0$, implying that $\theta$-type dominates the population as $t$ approaches infinity for any initial condition $(x_0, y_0)$. Hence, through endogenous matching, conditional cooperators can completely shut themselves from matching with non-cooperators, ensuring their prevalence in the long run. 

\subsection{Incomplete Information} \label{sec:incomplete}
In this section, we consider the case with incomplete information. That is, the agents' types are not observable. In this case, since the agents cannot identify the types of others, they cannot 
discriminate between whom they match with.
%match with those they prefer even they are given the opportunity to do so.
Hence, matching is uniformly random. We assume that the population state is common knowledge for the agents and the agents play a perfect Bayesian equilibrium (PBE) with one another after they are matched. First observe that in any PBE, a $\tau$-type agent would always choose her strictly dominant strategy $D$ as player 2, and a $\theta$-type agent as player 2 would choose $C$ upon observing that player 1 chooses $C$, and she would chooses $D$ upon observing that player 1 chooses $D$.  Lemma \ref{lemma:PBE} summarizes the choices made by player 1 in PBEs. We do not consider the possibility that the agents play mixed strategies. Hence, to break ties, we assume that the agents choose $C$ whenever they are indifferent between choosing $C$ and $D$.

\begin{lemma} \label{lemma:PBE}
Under incomplete information, 
\begin{itemize}
\item[(1)] When $\frac{x_t}{x_t+y_t} \ge \Delta_\theta:=\frac{\pi(D, D)-\pi(C, D)}{\pi(C, C)+\alpha-\pi(C, D)}$, a $\theta$-type player 1 chooses $C$ in PBE. Otherwise, she chooses $D$ in PBE. 
\item[(2)] When $\frac{x_t}{x_t+y_t} \ge \Delta_\tau:=\frac{\pi(D, D)-\pi(C, D)}{\pi(C, C)-\pi(C, D)}$, a $\tau$-type player 1 chooses $C$ in PBE. Otherwise, she chooses $D$ in PBE. 
\end{itemize}
\end{lemma}

\begin{figure}[!ht]%
\caption{Player 1's expected utility for each strategy \label{figure: incomplete information utilities}}
\centering
\includegraphics[scale=1]{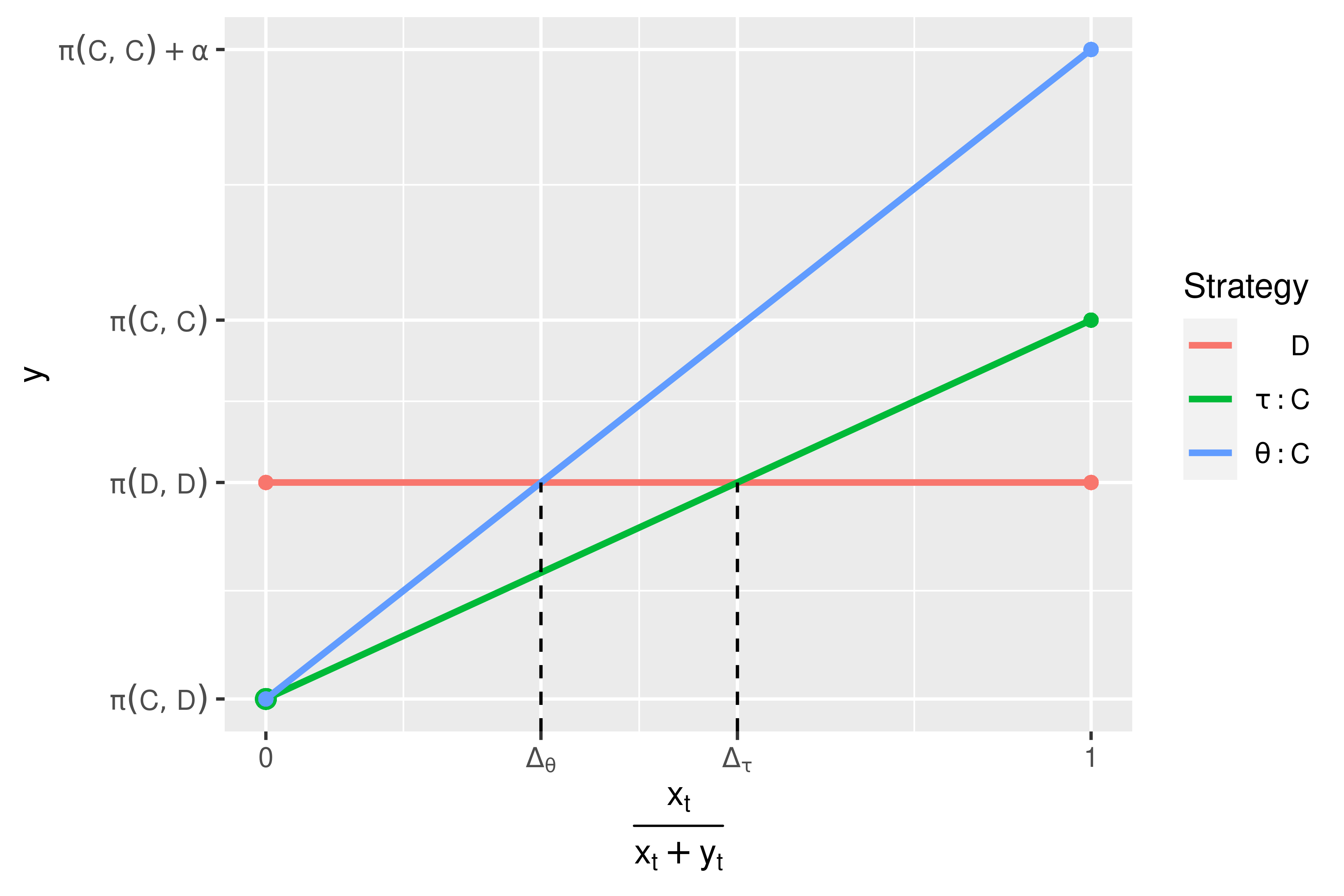}

\caption*{\footnotesize Note that it is always the case that $0 < \Delta_\theta < \Delta_\tau < 1$. Where $\Delta_X$ is the minimum proportion of conditional cooperators in the population required to make the expected utility of playing $C$ as the first player at least as high as the utility received from playing $D$ for player type $X \in \{\theta, \tau\}$.}

\end{figure}

Lemma \ref{lemma:PBE} which characterizes player 1's decision is illustrated by Figure \ref{figure: incomplete information utilities}. Given Lemma \ref{lemma:PBE}, we can characterize the long-run prediction of the evolutionary process of preferences:

\begin{proposition} \label{prop:incomplete}
Under incomplete information, a steady state population is characterized by $\frac{x_t}{x_t+y_t}\in (0, \Delta_\theta)$.
\end{proposition}

\begin{figure}[!ht]%
\caption{Evolution of Types under Incomplete Information \label{figure: incomplete information evolution}}

\begin{subfigure}{.55\linewidth}
\centering
\includegraphics[height=7cm]{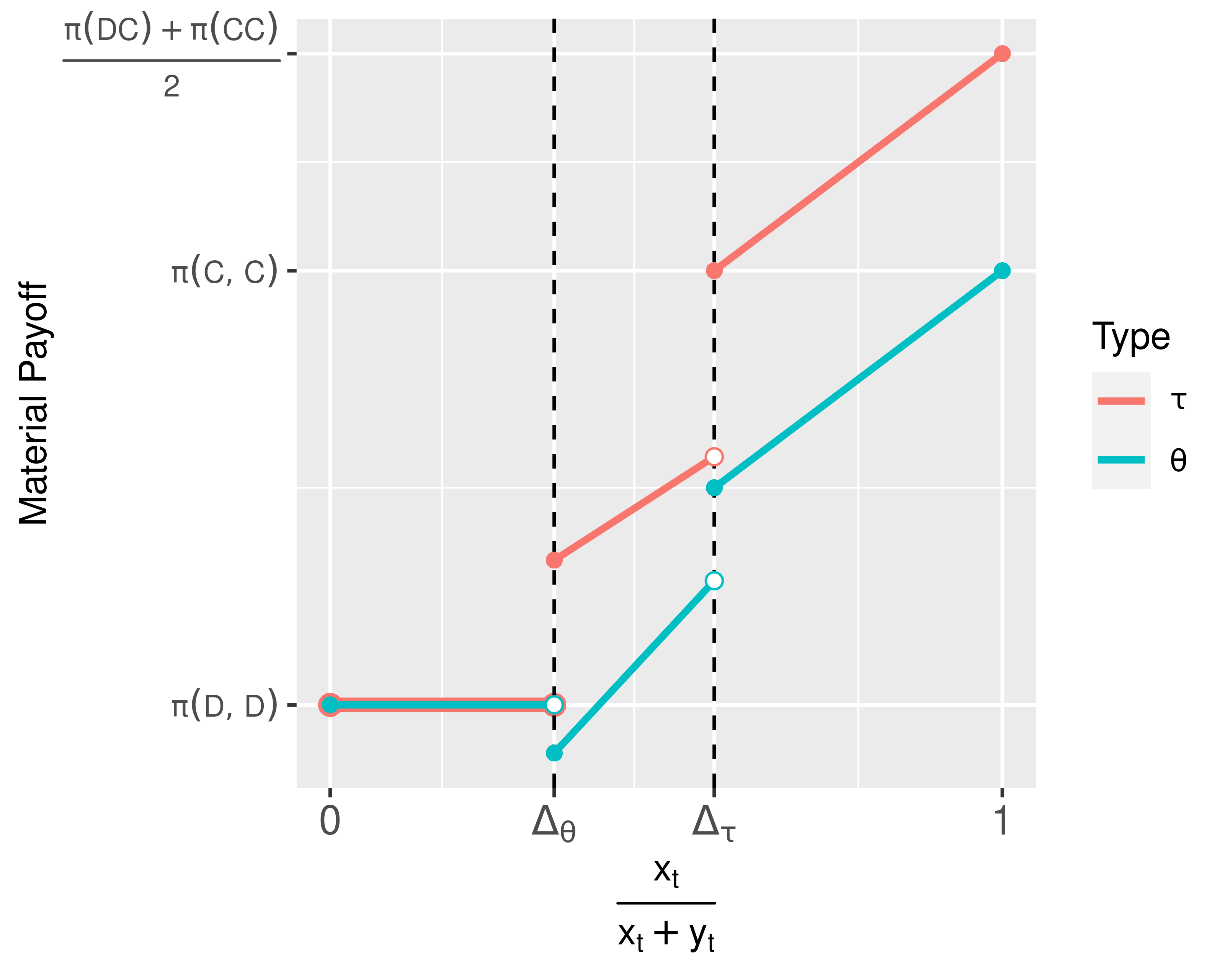}
\caption{Material Payoffs Realized for Each Type}
\label{figure: test1}
\end{subfigure}
\begin{subfigure}{.45\linewidth}
\centering
\includegraphics[height=7cm]{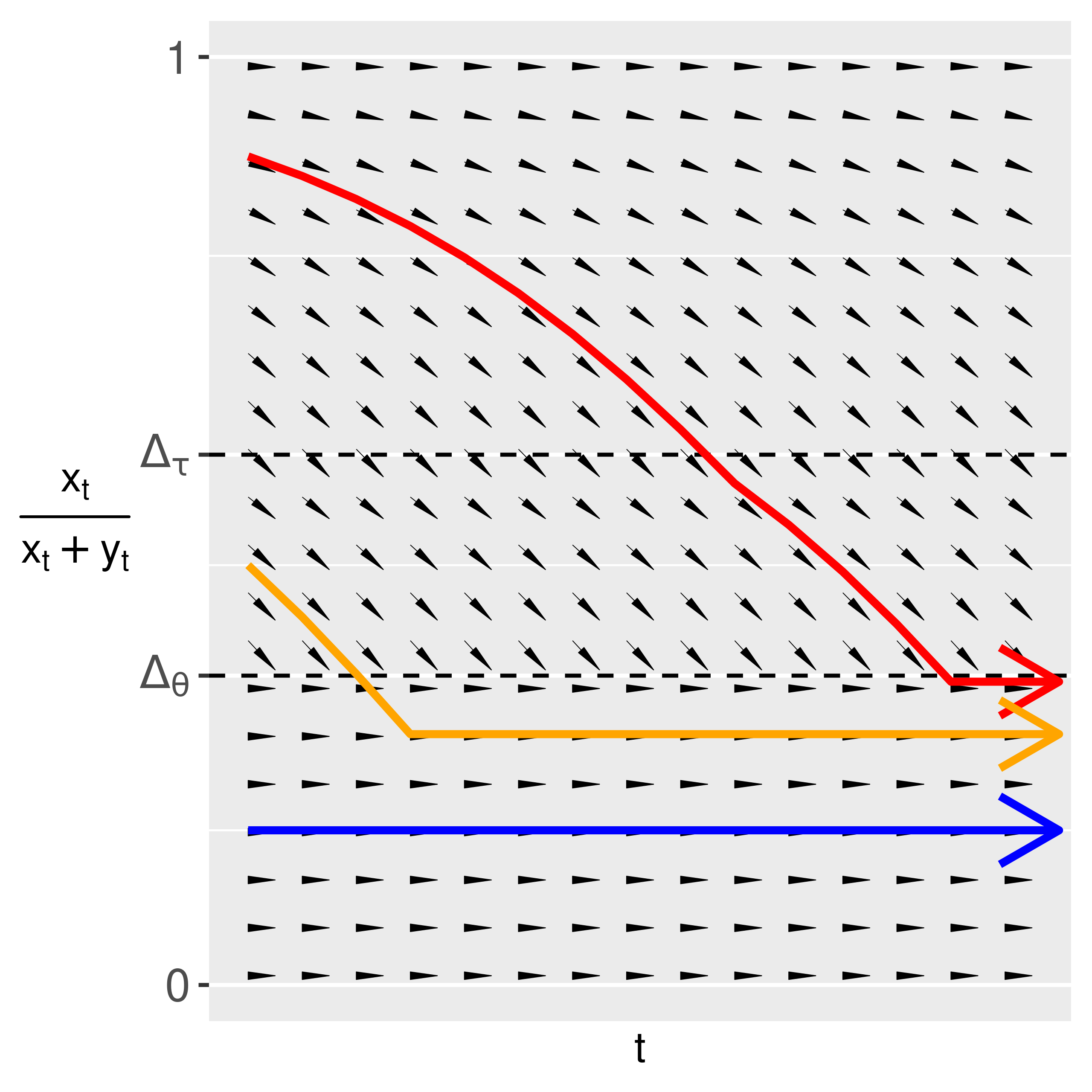}
\caption{Evolutionary Paths}
\label{figure: test1}
\end{subfigure}

\caption*{\footnotesize When $\frac{x_t}{x_t+y_t} \geq \Delta_\theta$ the average material payoffs are always greater for $\tau$ types than $\theta$ types as illustrated in plot (a). This results in the proportion of $\theta$ types in the population ($\frac{x_t}{x_t+y_t}$) decreasing until the level is below $\Delta_\theta$. This dynamic can be seen in plot (b). When $\frac{x_t}{x_t+y_t} < \Delta_\theta$ both types get the same average material payoff and so a steady state is realized.}

\end{figure}

The relationship between the average realized material payoffs by each type and the evolution of the population composition are depicted in graph (a) and (b) respectively in Figure \ref{figure: incomplete information evolution}.

Proposition \ref{prop:incomplete} shows that under incomplete information, the conditional cooperators can no longer dominate the entire society in the long run because they are not able to protect themselves from non-cooperators' exploitation. Nevertheless, they can still co-exist with the non-cooperators when their group size is sufficiently small so that they choose to defect as the non-cooperators. Note that the set $[0, \Delta_\theta)$ shrinks as $\alpha$ increases. In other words, a stronger psychological benefit of conditional cooperation makes the conditional cooperators more vulnerable in evolution. Note importantly, the results obtained in this section are merely the reminiscence of the results of \cite{guthkliemt1998RS}'s analysis on trust games.

\subsection{Institutional Screening} \label{sec:myopic}
In this section, we introduce an institution to the society. The institution is characterized by the \textbf{\textit{utilitarian}} social welfare function:
\begin{eqnarray}
W^{util}(x_t, y_t)=x_t\bar{U}_\theta(x_t, y_t)+y_t \bar{U}_\tau(x_t, y_t),
\end{eqnarray}
which is the total utility of the population in the current period. The institution aims to maximize the social welfare function by affecting the matching outcome. Since agents’ types are unobservable, the institution cannot directly match agents according to their types. Hence, the institution may want to implement certain screening scheme to isolate different types of agents if differentiating them increases social welfare. Note that we assume the institution is short sighted as it only cares about the social welfare of the population in the current period. We believe that this assumption is reasonable because policy cycles often have a much shorter duration than a single generation.

To differentiate different types of agents, the institution can set up a special zone that requires an entry fee. The entry fee should be set in way that only the conditional cooperators would enter. Alternatively, the institution can create a costly label such that only the conditional cooperators are willing to pay for having the label. If the screening scheme works, it would be common knowledge that all agents in the special zone or those who have the labels are conditional cooperators. Assume that the entry fee or the cost of label is collected as tax and redistributed to everyone evenly, so that it would not directly affect the social welfare. Observe that the unique stable matching given that all conditional cooperators are in the special zone or having the costly label is characterized by perfectly assortative matching since we are back to the complete information scenario.

Let the institution decision to implement a screening scheme be denoted by $z_t\in \{0, 1\}$ at time $t$. $z_t$ can be regarded as the policy variable of the institution. Rewrite $W^{util}(x_t, y_t)=W^{util}(x_t, y_t, z_t)$ to reflect that it is a function of $z_t$. We first ask the question whether the institution wants to implement a screening mechanism to segregate the conditional cooperators from the defectors. Then we examine how to set the entry fee for the special zone or the cost for the label such that it is incentive compatible that only the conditional cooperators would enter the special zone or purchase the label.

\begin{lemma} \label{lemma:optimal}
When $\frac{x_t}{x_t+y_t}<\Delta_\tau$, $z^*_t=1$. When $\frac{x_t}{x_t+y_t} \ge \Delta_\tau$, 1) if $\frac{x_t}{x_t+y_t} > \Delta_W := \frac{\pi(C, D)+\pi(D, C)-2\pi(D, D)}{\alpha}$, $z^*_t=1$; 2) if $\frac{x_t}{x_t+y_t} < \Delta_W$ , $z^*_t=0$; 3) if $\frac{x_t}{x_t+y_t} = \Delta_W$, $z^*_t=0$ or $1$.
%When $\frac{x_t}{x_t+y_t}<\Delta_\tau$, $z^*_t=1$. When $\frac{x_t}{x_t+y_t} \ge \Delta_\tau$, 1) if $\alpha >\frac{x_t+y_t}{x_t}(\pi(C, D)+\pi(D, C)-2\pi(D, D))$, $z^*_t=1$; 2) if $\alpha <\frac{x_t+y_t}{x_t}(\pi(C, D)+\pi(D, C)-2\pi(D, D))$, $z^*_t=0$; 3) if $\alpha = \frac{x_t+y_t}{x_t}(\pi(C, D)+\pi(D, C)-2\pi(D, D))$, $z^*_t=0$ or $1$.
\end{lemma}

\begin{figure}[!ht]%
\caption{Average Utility in the Population Under Varying Levels of $\Delta_W$  \label{figure: Screening Utilities}}
\centering
\includegraphics[width = \linewidth]{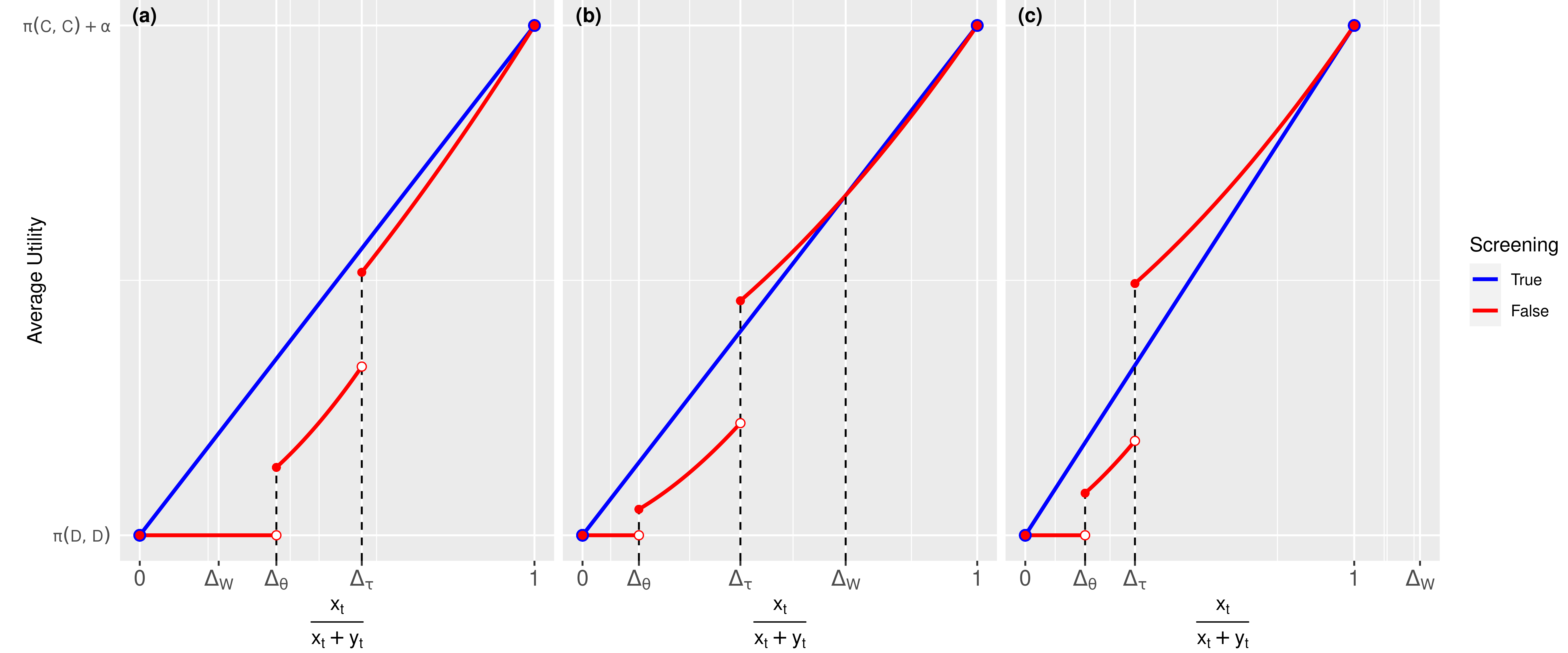}

\caption*{\footnotesize Presented are the three general cases for the position of $\Delta_W$. In (a), $\Delta_W \leq \Delta_\tau$ so screening yields a higher average utility for all type distributions than not screening. In (b), $\Delta_\tau < \Delta_W < 1$. In this case screening yields a higher average utility than not screening only when $\frac{x_t}{x_t+y_t} \in (0, \Delta_\tau) \cup (\Delta_W, 1)$. In (c), the case where $\Delta_W \geq 1$, screening yields a higher average utility only when  $\frac{x_t}{x_t+y_t} < \Delta_\tau$.}

\end{figure}

The institution's screening decision is depicted in Figure \ref{figure: Screening Utilities}. Let $\delta$ denote the entry fee for the special zone or the cost of label. Lemma \ref{lemma:IC1} characterizes the range of $\delta$ that would make the institution's screening mechanism incentive compatible. 

\begin{lemma}\label{lemma:IC1}
For any $\delta \in [\pi(D, C)-\pi(D, D), \pi(C, C)+\alpha-\pi(D, D)]$, $z^*_t=x_t$ can be implemented. 
\end{lemma}

Lemma \ref{lemma:IC1} shows that the institution is able to differentiate the conditional cooperators from the non-cooperators according to its goal stated in Lemma \ref{lemma:optimal}. 

Now, we can look at how institutional intervention would affect the evolution of preference types. We make two additional assumptions before we proceed with the analysis. First, since the end goal of the institution is to segregate the two types of agents, but not to maximize revenue (fees collected from the agents) we believe it is reasonable to assume that the institution would choose the minimum value of $\delta$, $\frac{1}{2}\pi(C, C)+\frac{1}{2}\pi(D, C)-\pi(D, D)$, if it wants to implement a screening mechanism. 
Second, we allow the price of the membership to the special zone or label to be discounted by parameter $p \in [0,1]$ for descendants of those already in the special zone or label holders. Hence, the cost for descendants of the special zone or label is only $\delta * (1-p)$. 
When $p = 1$, there is no cost for descendants, they all inherit the label or membership to the special zone. This is analogous to most cases of citizenship where it may be costly for parents to travel to a country and gain citizenship but once there, their progeny gain citizenship for free. When $p = 0$, each generation has to pay the full cost of membership. $p \in (0,1)$ means that descendants of the special zone or label have to pay some cost to maintain their status but not the full amount. Perhaps there are two costs for membership: an initiation fee or transaction cost and a maintenance fee. Such a setup can be seen from country club memberships to real estate. The descendants inherit the membership from their parents and only have to pay the ongoing maintenance fees. Those born without membership have to also pay the initiation fee. In this scenario, the value of $p$ is equal to the proportion of the initiation fee to the total cost of membership. Note that this is equivalent to a model where $p$ descendants inherit membership to the special zone or label. In this case, to maintain the screening scheme at generational $t>0$, the institution only needs to collect entry fee to the special zone or the cost of label from $x^*_t(1-p)$ amount of $\theta$-type agents.

\begin{proposition}\label{prop:institution}
Given the institutional intervention, $\tau$-type dominates the population as $t$ approaches infinity for any initial state $(x_0, y_0)$ if $p$ is sufficiently small. On the other hand, $\theta$-type dominates the population as $t$ approaches infinity for any initial state $(x_0, y_0)$ if $p$ and $\alpha$ are sufficiently large.
\end{proposition}

Recall that we assume that $\delta=\frac{1}{2}\pi(C, C)+\frac{1}{2}\pi(D, C)-\pi(D, D)$. Therefore, which type realizes higher material payoffs under screening depends on the relationship between $p$ and a threshold $\Delta_p := \frac{\pi(D, C)-\pi(C, C)}{\pi(C, C)+\pi(D, C)-2\pi(D, D)}$. In the proof, we show that the dynamic behavior of preference evolution is different across three scenarios: 1) $\Delta_W \geq1$, 2) $\Delta_W \leq \Delta_\tau$ and 3) $\Delta_W \in (\Delta_\tau, 1)$.

\begin{figure}[!ht]%
\caption{Material Payoff by Type under Institutional Screening when $\Delta_W \in (\Delta_\tau, 1)$ \label{figure: Screening Material Payoffs Case 2}}
\centering
\includegraphics[width = \linewidth]{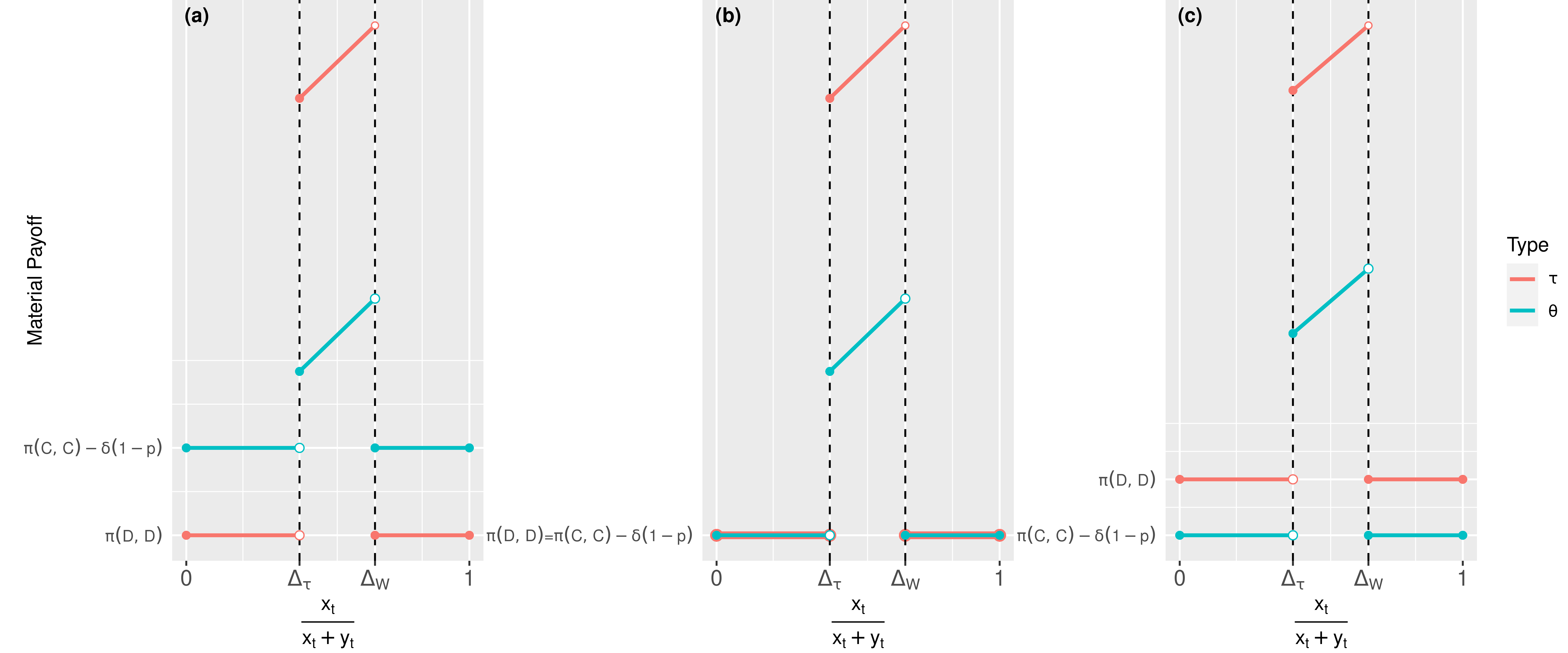}

\caption*{\footnotesize Between $\Delta_\tau$ and $\Delta_W$, there is no screening and the $\tau$ types receive a higher payoff. In (a), $p > \Delta_p$ so the $\theta$ types receive a higher material payoff than $\tau$ types under screening. In (b), $p = \Delta_p$. In this case, material payoffs are the same for both types under screening. In (c), the case where $p < \Delta_p$, $\tau$ types receive higher material payoffs than $\theta$ types under screening.}

\end{figure}

\begin{figure}[!ht]%
\caption{Evolutionary Paths under Institutional Screening when $\Delta_W \in (\Delta_\tau, 1)$  \label{figure: Screening Paths Case 2}}
\centering
\includegraphics[width = \linewidth]{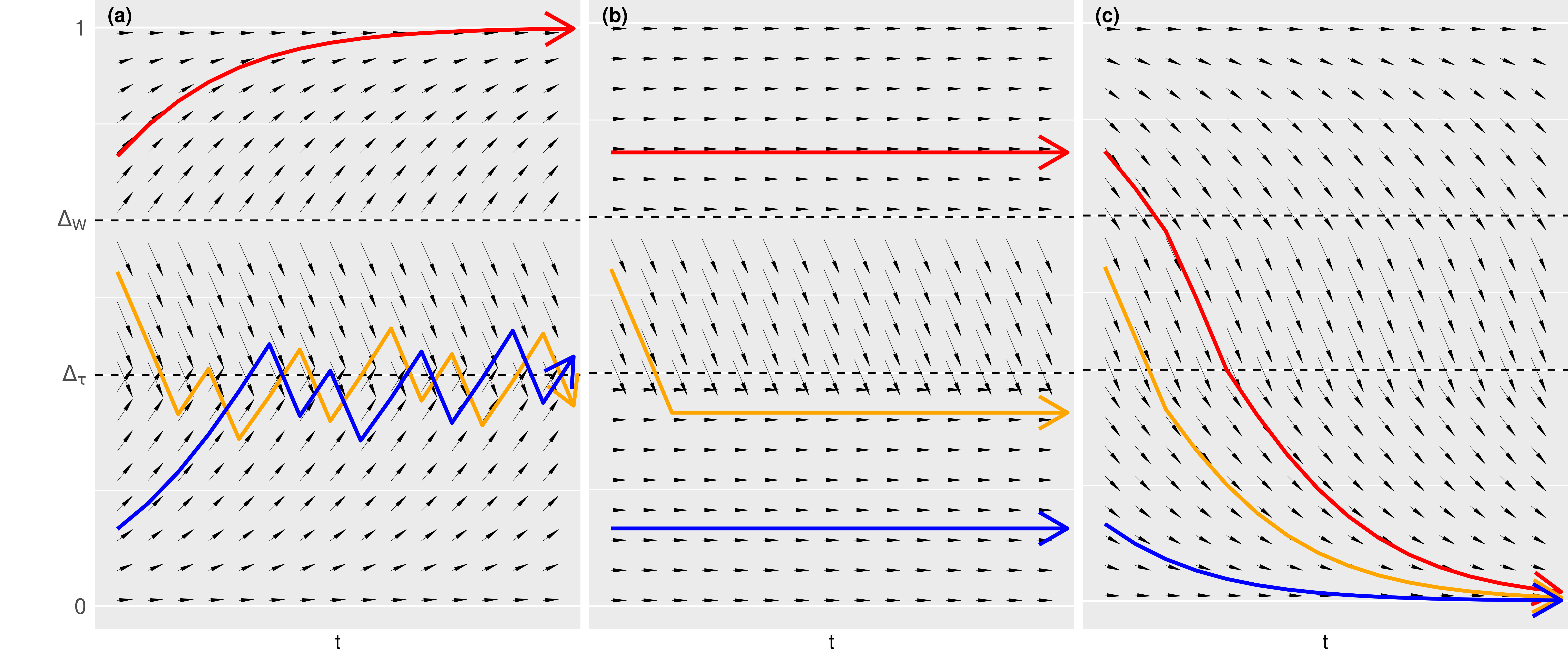}

\caption*{\footnotesize In (a), $p > \Delta_p$. If the proportion of $\theta$ types in the population is greater than $\Delta_W$ then $\theta$ types will evolve to dominate the population. Otherwise, the proportion of $\theta$ types in the population will oscillate around $\Delta_\tau$. In (b), since $p = \Delta_p$ any population proportion under screening is a steady state. If $\frac{x_t}{x_t+y_t} \in [\Delta_W, \Delta_\tau]$ then the institution will not implement screening until the proportion of $\theta$ types falls below $\Delta_\tau$. In (c), the case where $p < \Delta_p$, $\tau$ types receive higher material payoffs than $\theta$ types under both screening and no screening and will evolve to dominate the population.}

\end{figure}

\begin{figure}[!ht]%
\caption{Evolutionary Paths under Institutional Screening when $\Delta_W \leq \Delta_\tau$  \label{figure: Screening Paths Case 1}}
\centering
\includegraphics[width = \linewidth]{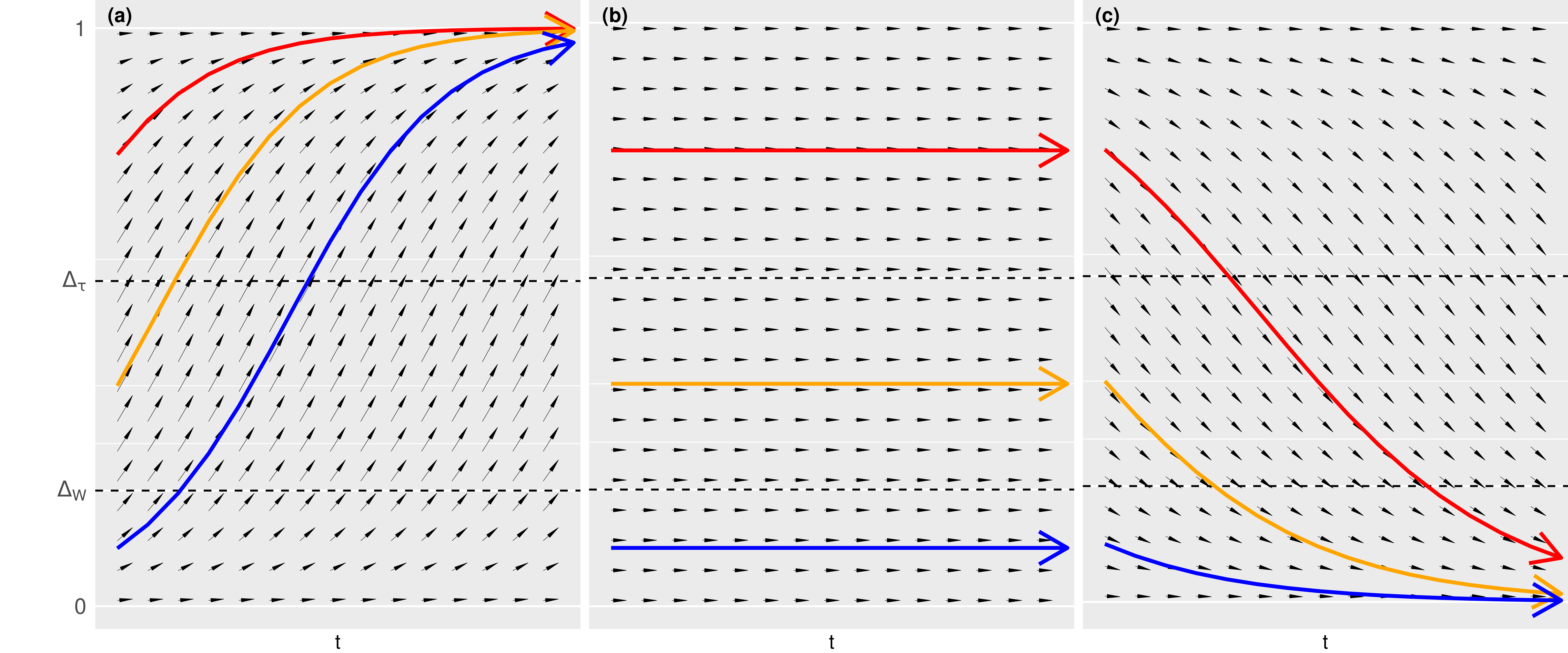}

\caption*{\footnotesize In (a), $p > \Delta_p$ and the $\theta$ types will evolve to dominate. In (b), since $p = \Delta_p$ any population proportion is a steady state. In (c), the case where $p < \Delta_p$, $\tau$ types will evolve to dominate the population.}
\end{figure}

\begin{figure}[!ht]%
\caption{Evolutionary Paths under Institutional Screening when $\Delta_W \geq 1$  \label{figure: Screening Paths Case 3}}
\centering
\includegraphics[width = \linewidth]{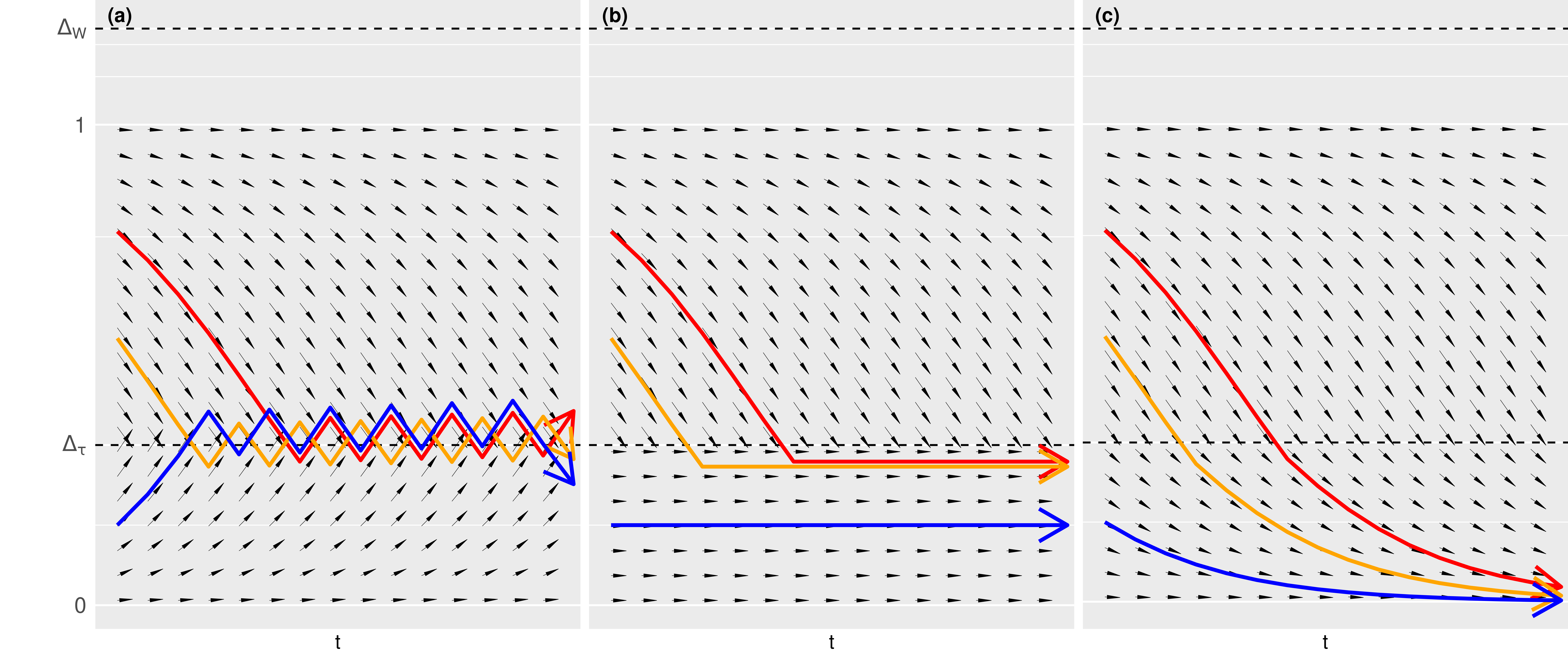}

\caption*{\footnotesize In (a), $p > \Delta_p$. There is no steady state and the proportion of $\theta$ types in the population will oscillate around $\Delta_\tau$. In (b), since $p = \Delta_p$ any population proportion under screening is a steady state. If $\frac{x_t}{x_t+y_t} \in [\Delta_\tau, 1]$ then the institution will not implement screening until the proportion of $\theta$ types falls below $\Delta_\tau$. In (c), the case where $p < \Delta_p$, $\tau$ types receive higher material payoffs than $\theta$ types under both screening and no screening and will evolve to dominate the population.}

\end{figure}

Illustrations of the average material payoffs for each type in the case where $\Delta_W \in (\Delta_\tau, 1)$ can be found in Figure \ref{figure: Screening Material Payoffs Case 2}. Figures of the average material payoffs for each type in the other cases are relegated to the appendix. A representation of the resulting evolutionary paths in each case can be found in Figures \ref{figure: Screening Paths Case 2}, \ref{figure: Screening Paths Case 1}, and \ref{figure: Screening Paths Case 3}.

Proposition  \ref{prop:institution} shows that whether conditional cooperators can prevail through evolution in the long run depends on two critical factors: 1) a sufficiently large psychological benefit of conditional cooperation $\alpha$ to make sure that the institution has a strong incentive to differentiate the conditional cooperators from the defectors, 2) a sufficient high probability of inheritance, which lowers the burden of future generations of conditional cooperators on paying the entry fee for the special zone  or the label cost. 

\section{Conclusion} \label{sec:conclusion}
We study the evolution of conditional cooperation under incomplete information by using the indirect evolutionary approach. We introduce an institution to the society, who can implement a screening scheme to differentiate the conditional cooperators from the non-cooperators. A screening scheme may benefit the conditional cooperators because it allows the conditional cooperators to match with and cooperate with other conditional cooperators. It is also desirable from the societal perspective in the short run as it can increase social welfare. In the long run, whether such a screening scheme can help to sustain cooperation through populating the society with conditional cooperators depends on several factors including the psychological benefit of conditional cooperation for the conditional cooperators and the probability of inheritance of the special zone membership or the label. 
 
In this paper, we consider institutional screening as a way to differentiate different agents in the population, which can be viewed as a centralized approach. On the other hand, costly signaling by \cite{Zahavi1975} and \cite{Spence1973} can serve as an alternative way from a decentralized perspective.\footnote{See \cite{Grafen1990}, \cite{MaynardSmithHarper1995}, \cite{Johnstone1997},  \cite{ZahaviZahavi1997}, \cite{Gintisetal2001} \cite{MaynardSmithHarper2003}, \cite{SearcyNowicki2005}, \cite{Getty2006}, \cite{Grose2011} and \cite{Szamado2012}, \cite{Hopkins2014AEJMICRO}, \cite{PrzepiorkaDiekmann2021}, \cite{HoldahlWu2023JEIC}, among many others, for works that apply the idea to explaining phenomena ranging from life sciences to social sciences and further discussions on the subject.} How signaling affects the evolution of conditional cooperation in the framework of the indirect evolutionary approach would be an interesting research question. 

\bibliographystyle{abbrvnat}
\bibliography{bib}

\begin{thebibliography}{39}
\providecommand{\natexlab}[1]{#1}
\providecommand{\url}[1]{\texttt{#1}}
\expandafter\ifx\csname urlstyle\endcsname\relax
  \providecommand{\doi}[1]{doi: #1}\else
  \providecommand{\doi}{doi: \begingroup \urlstyle{rm}\Url}\fi

\bibitem[Alger and Weibull(2019)]{AlgerWeibull2019ARE}
I.~Alger and J.~W. Weibull.
\newblock Evolutionary models of preference formation.
\newblock \emph{Annual Review of Economics}, 11:\penalty0 329--354, August
  2019.

\bibitem[Bisin and Verdier(2021)]{BisinVerdier2021WP}
A.~Bisin and T.~Verdier.
\newblock On the joint evolution of culture and institutions: Elites and civil
  society.
\newblock 2021.

\bibitem[Carvalho and Sacks(2021)]{CarvalhoSacks2021JPUBE}
J.-P. Carvalho and M.~Sacks.
\newblock The economics of religious communities.
\newblock \emph{Journal of Public Economics}, 201:\penalty0 104481, 2021.

\bibitem[David(1994)]{David1994}
P.~A. David.
\newblock Whyare institutions the `carriers of history'?: Path dependence and
  the evolution of conventions, organizations and institutions.
\newblock \emph{Structural Change and Economic Dynamics}, 5:\penalty0 205--220,
  1994.

\bibitem[Dufwenberg and Kirchsteiger(2004)]{DufwenbergKirchsteiger2004GEB}
M.~Dufwenberg and G.~Kirchsteiger.
\newblock A theory of sequential reciprocity.
\newblock \emph{Games and Economic Behavior}, 47:\penalty0 268--298, 2004.

\bibitem[Echenique et~al.(2013)Echenique, Lee, Shum, and
  Yenmez]{Echeniqueetal2013ECMA}
F.~Echenique, S.~Lee, M.~Shum, and B.~M. Yenmez.
\newblock The revealed preference theory of stable and extremal stable
  mathcings.
\newblock \emph{Econometrica}, 81\penalty0 (1):\penalty0 153--171, 2013.

\bibitem[Fischbacher and G\"{a}chter(2010)]{FG2010AER}
U.~Fischbacher and S.~G\"{a}chter.
\newblock Social preferences, beliefs, and the dynamics of free riding in
  public goods experiments.
\newblock \emph{American Economic Review}, 100\penalty0 (1):\penalty0 541--556,
  2010.

\bibitem[Fischbacher et~al.(2001)Fischbacher, G\"{a}chter, and Fehr]{FGF2001EL}
U.~Fischbacher, S.~G\"{a}chter, and E.~Fehr.
\newblock Are people conditionally cooperative? evidence from a public goods
  experiment.
\newblock \emph{Economics Letters}, 71\penalty0 (3):\penalty0 397--404, 2001.

\bibitem[Gale and Shapley(1962)]{GaleShapley1962AMM}
D.~Gale and L.~S. Shapley.
\newblock College admissions and the stability of marriage.
\newblock \emph{The American Mathematical Monthly}, 69\penalty0 (1):\penalty0
  9--15, January 1962.

\bibitem[Getty(2006)]{Getty2006}
T.~Getty.
\newblock Sexually selected signals are not similar to sports handicaps.
\newblock \emph{Trends in ecology and evolution}, 21:\penalty0 83--88, 2006.

\bibitem[Gintis et~al.(2001)Gintis, Alden~Smith, and Bowles]{Gintisetal2001}
H.~Gintis, E.~Alden~Smith, and S.~Bowles.
\newblock Costly signaling and cooperation.
\newblock \emph{Journal of Theoretical Biology}, 213:\penalty0 103--119, 2001.

\bibitem[Grafen(1990)]{Grafen1990}
A.~Grafen.
\newblock Biological signals as handicaps.
\newblock \emph{Journal of Theoretical Biology}, 144:\penalty0 517--546, 1990.

\bibitem[Grief(2006)]{Grief2006}
A.~Grief.
\newblock \emph{Institutions and the Path to the Modern Economy: Lessons from
  Medieval Trade}.
\newblock Cambridge University Press, New York, 2006.

\bibitem[Grose(2011)]{Grose2011}
J.~Grose.
\newblock Modelling and the fall and rise of the handicap principle.
\newblock \emph{Biology \& Philosophy}, 26:\penalty0 677--696, 2011.

\bibitem[G\"{u}th(1995)]{guth1995ijgt}
W.~G\"{u}th.
\newblock An evolutionary approach to explaining cooperative behavior by
  reciprocal incentives.
\newblock \emph{International Journal of Game Theory}, 24:\penalty0 323--344,
  1995.

\bibitem[G\"{u}th and Kliemt(1998)]{guthkliemt1998RS}
W.~G\"{u}th and H.~Kliemt.
\newblock Indirect evolutionary approach: Bridging the gap between rationality
  and adaptation.
\newblock \emph{Rationality and Society}, 10\penalty0 (3):\penalty0 377--399,
  1998.

\bibitem[G\"{u}th and Yaari(1992)]{guthyaari1992}
W.~G\"{u}th and M.~Yaari.
\newblock An evolutionary approach to explaining reciprocal behavior in a
  simple strategic game.
\newblock In U.~Witt, editor, \emph{Approaches to Evolutionary Economics}.
  University of Michigan Press, 1992.

\bibitem[Hiller et~al.(2022)Hiller, Wu, and Zhang]{Hilleratal2022WP}
V.~Hiller, J.~Wu, and H.~Zhang.
\newblock Heterophily, stable matching, and intergenerational transmission in
  cultural evolution.
\newblock 2022.

\bibitem[Holdahl and Wu(2023)]{HoldahlWu2023JEIC}
E.~Holdahl and J.~Wu.
\newblock Conflicts, assortative matching, and the evolution of signaling
  norms.
\newblock \emph{Journal of Economic Interaction and Coordination}, pages
  936--941, 2023.

\bibitem[Hopkins(2014)]{Hopkins2014AEJMICRO}
E.~Hopkins.
\newblock Competitive altruism, mentalizing and signaling.
\newblock \emph{American Economic Journal: Microeconomics}, 2656\penalty0
  (4):\penalty0 272--292, 2014.

\bibitem[Iannaccone(1992)]{Iannaccone1992JPE}
L.~R. Iannaccone.
\newblock Sacrifice and stigma: Reducing free-riding in cults, communes, and
  other collectives.
\newblock \emph{Journal of Political Economy}, 100\penalty0 (2):\penalty0
  271--291, 1992.

\bibitem[Johnstone(1997)]{Johnstone1997}
R.~Johnstone.
\newblock The evolution of animal signals.
\newblock In J.~Krebs and N.~Davies, editors, \emph{Behavioral ecology, an
  evolutionary approach}, pages 465--485. Blackwell Scientific Publications,
  Oxford, 4 edition, 1997.

\bibitem[Maynard~Smith and Harper(1995)]{MaynardSmithHarper1995}
J.~Maynard~Smith and D.~Harper.
\newblock Animal signals: models and terminology.
\newblock \emph{Journal of Theoretical Biology}, 177:\penalty0 305--311, 1995.

\bibitem[Maynard~Smith and Harper(2003)]{MaynardSmithHarper2003}
J.~Maynard~Smith and D.~Harper.
\newblock \emph{Animal Signals}.
\newblock Oxford University Press, Oxford, 2003.

\bibitem[North(1981)]{North1981}
D.~North.
\newblock \emph{Structure and Change in Economic History}.
\newblock Norton, New York, 1981.

\bibitem[North(1990)]{North1990}
D.~North.
\newblock \emph{Institutions, Institutional Change and Economic Performance}.
\newblock Cambridge University Press, New York, 1990.

\bibitem[Ostrom(2000)]{Ostrom2000JEP}
E.~Ostrom.
\newblock Collective action and the evolution of social norms.
\newblock \emph{Journal of Economic Perspectives}, 14\penalty0 (3):\penalty0
  137--158, 2000.

\bibitem[Przepiorka and Diekmann(2021)]{PrzepiorkaDiekmann2021}
W.~Przepiorka and A.~Diekmann.
\newblock Parochial cooperation and the emergence of signalling norms.
\newblock \emph{Philosophical Transactions of the Royal Society B: Biological
  Sciences}, 376:\penalty0 20200294, 2021.

\bibitem[Rabin(1993)]{rabin1993aer}
M.~Rabin.
\newblock Incorporating fairness into game theory and economics.
\newblock \emph{American Economic Review}, 83:\penalty0 1281--1302, 1993.

\bibitem[Rigos and Nax(2016)]{RigosNax2016JTB}
A.~Rigos and H.~Nax.
\newblock Assortativity evolving from social dilemmas.
\newblock \emph{Journal of Theoretical Biology}, 395:\penalty0 194--203, 2016.

\bibitem[Searcy and Nowicki(2005)]{SearcyNowicki2005}
W.~Searcy and S.~Nowicki.
\newblock \emph{The evolution of animal communication. Reliability and
  deception in signalling systems}.
\newblock Princeton University Press, Princeton, 2005.

\bibitem[Spence(1973)]{Spence1973}
A.~Spence.
\newblock Job market signaling.
\newblock \emph{Quarterly Journal of Economics}, 87:\penalty0 355--374, 1973.

\bibitem[Sz\'{a}mad\'{o}(2012)]{Szamado2012}
S.~Sz\'{a}mad\'{o}.
\newblock The rise and fall of the handicap principle: a commentary on the
  ``modelling and the fall and rise of the handicap principle.
\newblock \emph{Biology \& Philosophy}, 27:\penalty0 279--286, 2012.

\bibitem[Wang and Wu(2022)]{WangWu2022WP}
Z.~Wang and J.~Wu.
\newblock Preference evolution under stable matching.
\newblock 2022.

\bibitem[Wu(2016)]{Wu2016EB}
J.~Wu.
\newblock Evolving assortativity and social conventions.
\newblock \emph{Economics Bulletin}, 36:\penalty0 936--941, 2016.

\bibitem[Wu(2021)]{Wu2021RED}
J.~Wu.
\newblock Matching markets and cultural selection.
\newblock \emph{Review of Economic Design}, 25:\penalty0 267--288, 2021.

\bibitem[Wu(2022)]{Wu2022WP}
J.~Wu.
\newblock Institutions, assortative matching, and cultural evolution.
\newblock 2022.

\bibitem[Zahavi(1975)]{Zahavi1975}
A.~Zahavi.
\newblock Mate selection: a selection for a handicap.
\newblock \emph{Journal of Theoretical Biology}, 53:\penalty0 205--214, 1975.

\bibitem[Zahavi and Zahavi(1997)]{ZahaviZahavi1997}
A.~Zahavi and A.~Zahavi.
\newblock \emph{The handicap principle}.
\newblock Oxford University Press, Oxford, 1997.

\end{thebibliography}

\pagebreak

\section*{Appendix}

\subsection*{A: Additional Figures}

\begin{figure}[!ht]%
\caption{Material Payoff by Type under Institutional Screening when $\Delta_W \leq \Delta_\tau$ \label{figure: Screening Material Payoffs Case 1}}
\centering
\includegraphics[width = \linewidth]{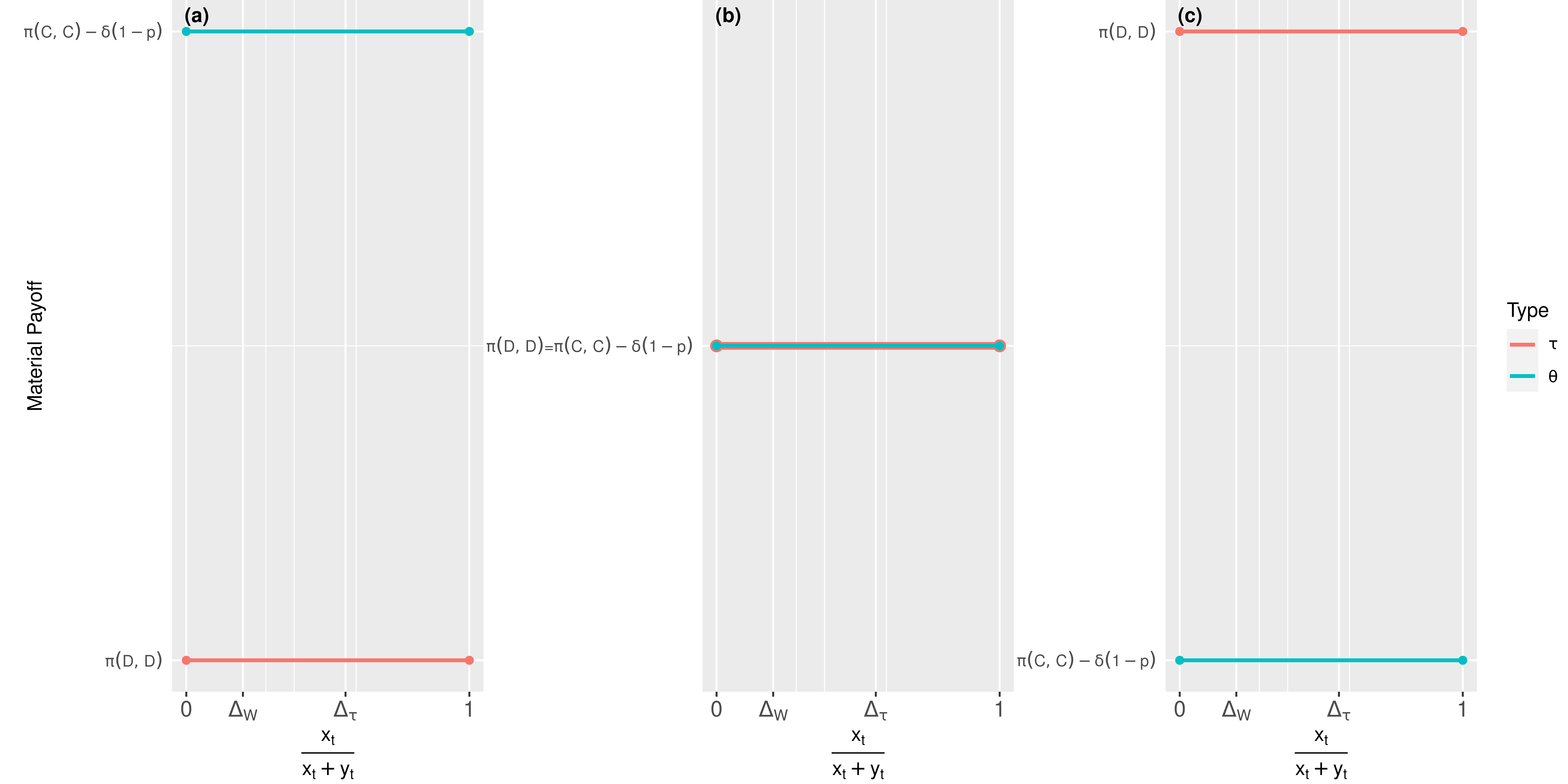}

\caption*{\footnotesize In this case there is always screening implemented. In (a), $p > \Delta_p$ so the $\theta$ types receive a higher material payoff than $\tau$ types under screening. In (b), $p = \Delta_p$. In this case, material payoffs are the same for both types under screening. In (c), the case where $p < \Delta_p$, $\tau$ types receive higher material payoffs than $\theta$ types under screening.}

\end{figure}

\begin{figure}[!ht]%
\caption{Material Payoff by Type under Institutional Screening when $\Delta_W \geq 1$ \label{figure: Screening Material Payoffs Case 3}}
\centering
\includegraphics[width = \linewidth]{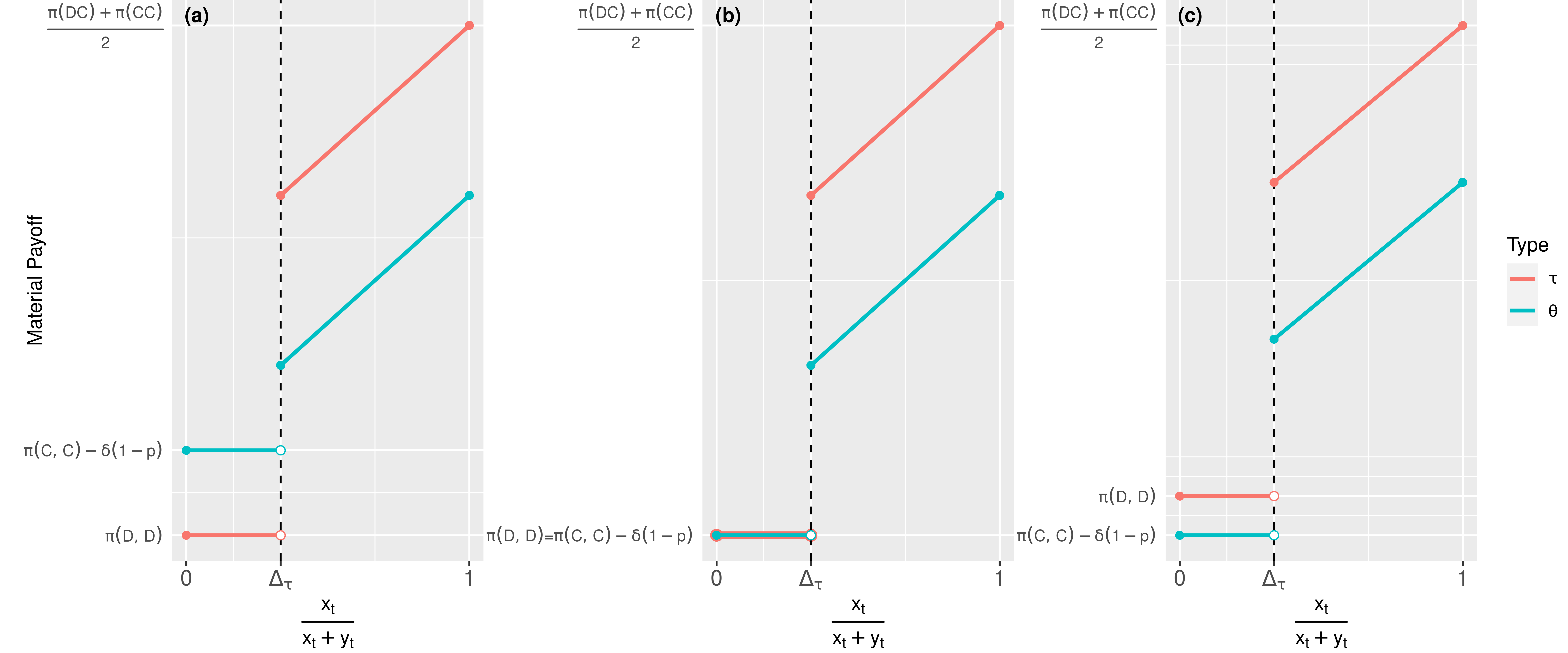}

\caption*{\footnotesize In this case screening implemented only when $\frac{x_t}{x_t+y_t} < \Delta_\tau$. Under no screening the $\tau$ types always recieve a higher material payoff than the $\theta$ types. In (a), $p > \Delta_p$ so the $\theta$ types receive a higher material payoff than $\tau$ types under screening. In (b), $p = \Delta_p$. In this case, material payoffs are the same for both types under screening. In (c), the case where $p < \Delta_p$, $\tau$ types receive higher material payoffs than $\theta$ types under screening.}

\end{figure}

\pagebreak

\subsection*{B: Proofs}

\noindent \textbf{Proof of Lemma \ref{lemma:PAM}:} First, we prove that it is a stable matching. The only potential blocking pair is $(\theta, \tau)$. However, since $K(\theta, \tau)=\frac{1}{2} \pi(D, D)+\frac{1}{2}(\pi(C, C)+\alpha)<K(\theta, \theta)=\pi(C, C)+\alpha$, it is not a blocking pair. Second, we prove that it is unique. Suppose there exists another stable matching such that $\mu_{\theta, \tau}(x_t, y_t)>0$, then $(\theta, \theta)$ is a blocking pair because $K(\theta, \tau)<K(\theta, \theta)$, which is a contradiction. \textit{Q.E.D.}
\\

\noindent \textbf{Proof of Lemma \ref{lemma:PBE}:}
Given what player 2 would choose in the second stage of the game, a $\theta$-type player 1's expected utility of choosing $C$ is $\frac{x_t}{x_t+y_t}(\pi(C, C)+\alpha)+\frac{y_t}{x_t+y_t}\pi(C, D)$; and her expected utility of choosing $D$ is $\pi(D, D)$. Hence, she chooses $C$ if and only if $\frac{x_t}{x_t+y_t} \ge \Delta_\theta=\frac{\pi(D, D)-\pi(C, D)}{\pi(C, C)+\alpha-\pi(C, D)}$. A $\tau$-type player 1's expected utility of choosing $C$ is $\frac{x_t}{x_t+y_t}\pi(C, C)+\frac{y_t}{x_t+y_t}\pi(C, D)$; and her expected utility of choosing $D$ is $\pi(D, D)$. Hence, she chooses $C$ if and only if $\frac{x_t}{x_t+y_t} \ge \Delta_\tau=\frac{\pi(D, D)-\pi(C, D)}{\pi(C, C)-\pi(C, D)}$. \textit{Q.E.D}
\\

\noindent \textbf{Proof of Proposition \ref{prop:incomplete}:}
When $\frac{x_t}{x_t+y_t}\ge \Delta_\tau$, the average material payoffs of the two types are given by 
\begin{eqnarray}
&&F_\theta(x_t, y_t)=\frac{x_t}{x_t+y_t}\pi(C, C)+\frac{y_t}{x_t+y_t}(\frac{1}{2}\pi(C, D)+\frac{1}{2}\pi(C, C)), \\
&& F_\tau(x_t, y_t)=\frac{x_t}{x_t+y_t}(\frac{1}{2}\pi(C, C)+\frac{1}{2}\pi(D, C))+\frac{y_t}{x_t+y_t}(\frac{1}{2}\pi(C, D)+\frac{1}{2}\pi(D, C)).
\end{eqnarray}
$F_\theta(x_t, y_t)-F_\tau(x_t, y_t)=-\frac{1}{2}(\pi(D,C)-\pi(C, C))<0$, which tells us that $\frac{x_t}{x_t+y_t}$ is decreasing in $t$. 

When $\Delta_\theta \le \frac{x_t}{x_t+y_t}< \Delta_\tau$, the average material payoffs of the two types are given by 
\begin{eqnarray}
&&F_\theta(x_t, y_t)=\frac{x_t}{x_t+y_t}\pi(C, C)+\frac{y_t}{x_t+y_t}(\frac{1}{2}\pi(C, D)+\frac{1}{2}\pi(D, D)), \\
&& F_\tau(x_t, y_t)=\frac{x_t}{x_t+y_t}(\frac{1}{2}\pi(D, D)+\frac{1}{2}\pi(D, C))+\frac{y_t}{x_t+y_t}\pi(D, D).
\end{eqnarray}
$F_\theta(x_t, y_t)>F_\tau(x_t, y_t)$ is equivalent to $\frac{x_t}{x_t+y_t}>\frac{\pi(D, D)-\pi(C, D)}{2\pi(C, C)-\pi(C, D)-\pi(D, C)}$. However, $\frac{\pi(D, D)-\pi(C, D)}{2\pi(C, C)-\pi(C, D)-\pi(D, C)}>\Delta_\tau$. Hence, $F_\theta(x_t, y_t)<F_\tau(x_t, y_t)$ when $\frac{x_t}{x_t+y_t} \in [\Delta_\theta, \Delta_\tau)$, which tells us that  $\frac{x_t}{x_t+y_t}$ is decreasing in $t$. 

When $ \frac{x_t}{x_t+y_t}<\Delta_\theta$, we must have $F_\theta(x_t, y_t)=F_\tau(x_t, y_t)=\pi(D, D)$. Hence, $\frac{x_t}{x_t+y_t}$ remains a constant and the population reaches a steady state. \textit{Q.E.D.}
\\

\noindent \textbf{Proof of Lemma \ref{lemma:optimal}:}
If the institution implements a screening scheme, we must have $W^{util}(x_t, y_t, 1)=x_t(\pi(C, C)+\alpha)+y_t\pi(D, D)$.

If the institution does not implement a screening mechanism, then we calculate the social welfare according to Lemma \ref{lemma:PBE}. First, suppose $\frac{x_t}{x_t+y_t} < \Delta_\theta$. In this case, we have $W^{util}(x_t, y_t, 0)=(x_t+y_t)\pi(D, D)<W^{util}(x_t, y_t, 1)$. Hence, $z^*_t=1$.

Second, suppose $\frac{x_t}{x_t+y_t} \in [\Delta_\theta, \Delta_\tau)$. In this case, we have
\begin{eqnarray}
W^{util}(x_t, y_t, 0)&=&x_t\big(\frac{x_t}{x_t+y_t}(\pi(C, C)+\alpha)+\frac{y_t}{x_t+y_t}(\frac{1}{2}\pi(C, D)+\frac{1}{2}\pi(D, D)\big)\nonumber \\
&&+ y_t\big(\frac{x_t}{x_t+y_t}(\frac{1}{2}\pi(D, D)+\frac{1}{2}\pi(D, C))+\frac{y_t}{x_t+y_t}\pi(D, D)\big).
\end{eqnarray} 
Simple calculation yields:
\begin{eqnarray}
W^{util}(x_t, y,1)-W^{util}(x_t, y_t, 0)=\frac{x_ty_t}{x_t+y_t}\big((\pi(C, C)+\alpha)-\frac{1}{2}\pi(C, D)-\frac{1}{2}\pi(D, C)\big)>0.
\end{eqnarray} 
Hence, $z^*_t=1$.

Suppose $\frac{x_t}{x_t+y_t} \geq\Delta_\tau$, In this case, we have
\begin{eqnarray}
&&W^{util}(x_t, y_t, 0)=x_t\big(\frac{x_t}{x_t+y_t}(\pi(C, C)+\alpha)+\frac{y_t}{x_t+y_t}(\frac{1}{2}\pi(C, D)+\frac{1}{2}(\pi(C, C)+\alpha))\big) \nonumber \\
&&+y_t\big(\frac{x_t}{x_t+y_t}(\frac{1}{2}\pi(C, C)+\frac{1}{2}\pi(D, C))+\frac{y_t}{x_t+y_t}(\frac{1}{2}\pi(C, D)+\frac{1}{2}\pi(D, C))\big).
\end{eqnarray} 
Simple calculation yields:
\begin{eqnarray}
W^{util}(x_t, y,1)-W^{util}(x_t, y_t, 0)=y_t(\frac{x_t}{x_t+y_t}\frac{1}{2}\alpha+\pi(D, D)-\frac{1}{2}(\pi(C, D)+\pi(D, C)).
\end{eqnarray}
which can be positive or negative depending on the comparison between $\frac{x_t}{x_t+y_t}$ and $\frac{\pi(C, D)+\pi(D, C)-2\pi(D, D)}{\alpha} = \Delta_W$. In sum, when $\frac{x_t}{x_t+y_t}<\Delta_\tau$, the institution always wants to implement a screening scheme. When $\frac{x_t}{x_t+y_t} \ge \Delta_\tau$, whether the institution prefers a screening mechanism depends on the value of $\Delta_W$.
%which can be positive or negative depending on the comparison between $\alpha$ and $\frac{x_t+y_t}{x_t}(\pi(C, D)+\pi(D, C)-2\pi(D, D))$. In sum, when $\frac{x_t}{x_t+y_t}<\Delta_\tau$, the institution always wants to implement a screening scheme. When $\frac{x_t}{x_t+y_t} \ge \Delta_\tau$, whether the institution prefers a screening mechanism depends on the value of $\alpha$.  
\textit{Q.E.D.}
\\

\noindent \textbf{Proof of Lemma \ref{lemma:IC1}:}
From an ex-ante perspective, given that the agents expect that all conditional cooperators are in the special zone or with the label, while all non-cooperators are outside the special zone or without the label and the unique stable matching is perfectly assortative, we need to ensure that the conditional cooperators have the incentive to pay the entry fee or the cost of label, while the non-cooperators do not. Hence, two conditions need to be satisfied at the same time:
\begin{eqnarray}
&& \pi(C, C)+\alpha-\delta \ge \pi(D, D); \label{IC1}\\
&& \frac{1}{2}\pi(C, C)+\frac{1}{2}\pi(D, C)-\delta \le \pi(D, D). \label{IC2}
\end{eqnarray}
They imply that $\delta \in [\frac{1}{2}\pi(C, C)+\frac{1}{2}\pi(D, C)-\pi(D, D), \pi(C, C)+\alpha-\pi(D, D)]$. Since $\pi(C, C)+\alpha>\pi(D, C)$ by assumption, the set is non-empty. \textit{Q.E.D.} 
\\

\noindent \textbf{Proof of Proposition \ref{prop:institution}:}

i) Suppose $\alpha \geq \frac{1}{\Delta_\tau}(\pi(C, D)+\pi(D, C)-2\pi(D, D))$. Thus, $\Delta_W \leq \Delta_\tau$ which means $z^*_{t}=1$ for any $(x_t, y_t)$. Hence,
%i) Suppose $\alpha > \frac{1}{\Delta_\tau}(\pi(C, D)+\pi(D, C)-2\pi(D, D))$, then $z^*_{t}=1$ for any $(x_t, y_t)$. Hence, 
\begin{eqnarray}
F_\theta(x_t, y_t)-F_\tau(x_t, y_t)=\pi(C, C)-\delta(1-p)-\pi(D, D). 
\end{eqnarray}

When $p>\Delta_p$, we must have $F_\theta(x_t, y_t)-F_\tau(x_t, y_t)>0$ for any $(x_t, y_t)$, implying that $\theta$-type dominates the population as $t$ approaches infinity for any initial condition $(x_0, y_0)$. When $p<\Delta_p$, $F_\theta(x_t, y_t)-F_\tau(x_t, y_t)<0$ for any $(x_t, y_t)$, implying that $\tau$-type dominates the population as $t$ approaches infinity for any initial condition $(x_0, y_0)$. When $p=\Delta_p$, $F_\theta(x_t, y_t)-F_\tau(x_t, y_t)=0$ for any $(x_t, y_t)$, implying that any state is a steady state. 

ii) Suppose $\alpha \leq \pi(C, D)+\pi(D, C)-2\pi(D, D)$. Then $\Delta_W \geq1$ and $z^*_{t}=1$ for any $\frac{x_t}{x_t+y_t}<\Delta_\tau$, and $z^*_t=0$ for any $\frac{x_t}{x_t+y_t}\ge \Delta_\tau$. This implies that $F_\theta(x_t, y_t)-F_\tau(x_t, y_t)<0$ for any $\frac{x_t}{x_t+y_t}\ge \Delta_\tau$. When $p>\Delta_p$, we must have $F_\theta(x_t, y_t)-F_\tau(x_t, y_t)>0$ for any $\frac{x_t}{x_t+y_t}<\Delta_\tau$. Hence, there is no steady state in this case. When $p<\Delta_p$, $F_\theta(x_t, y_t)-F_\tau(x_t, y_t)<0$ for any $(x_t, y_t)$, implying that $\tau$-type dominates the population as $t$ approaches infinity for any initial condition $(x_0, y_0)$. When $p=\Delta_p$, $F_\theta(x_t, y_t)-F_\tau(x_t, y_t)=0$  for any $\frac{x_t}{x_t+y_t}\in (0, \Delta_\tau)$, implying that any state such that $\frac{x_t}{x_t+y_t} \in (0, \Delta_\tau)$ is a steady state.
%ii) Suppose $\alpha<(\pi(C, D)+\pi(D, C)-2\pi(D, D))$, then $z^*_{t}=1$ for any $\frac{x_t}{x_t+y_t}<\Delta_\tau$, and $z^*_t=0$ for any $\frac{x_t}{x_t+y_t}\ge \Delta_\tau$. This implies that $F_\theta(x_t, y_t)-F_\tau(x_t, y_t)<0$ for any $\frac{x_t}{x_t+y_t}\ge \Delta_\tau$. When $p>\Delta_p$, we must have $F_\theta(x_t, y_t)-F_\tau(x_t, y_t)>0$ for any $\frac{x_t}{x_t+y_t}<\Delta_\tau$. Hence, there is no steady state in this case. When $p<\Delta_p$, $F_\theta(x_t, y_t)-F_\tau(x_t, y_t)<0$ for any $(x_t, y_t)$, implying that $\tau$-type dominates the population as $t$ approaches infinity for any initial condition $(x_0, y_0)$. When $p=\Delta_p$, $F_\theta(x_t, y_t)-F_\tau(x_t, y_t)=0$  for any $\frac{x_t}{x_t+y_t}\in (0, \Delta_\tau)$, implying that any state such that $\frac{x_t}{x_t+y_t} \in (0, \Delta_\tau)$ is a steady state. 

iii) Suppose $\alpha\in (\pi(C, D)+\pi(D, C)-2\pi(D, D), \frac{1}{\Delta_\tau}(\pi(C, D)+\pi(D, C)-2\pi(D, D)))$. Thus, $\Delta_W \in (\Delta_\tau, 1)$ such that for any $\frac{x_t}{x_t+y_t} \in (\Delta_W, 1)$, $z^*_{t}=1$, for any $\frac{x_t}{x_t+y_t} \in (\Delta_\tau, \Delta_W)$, $z^*_{t}=0$, and for $\frac{x_t}{x_t+y_t} = \Delta_W$, $z^*_{t}=0$ or $1$. Also, we still have $z^*_{t}=1$ for any $\frac{x_t}{x_t+y_t}<\Delta_\tau$. When $p>\Delta_p$, we must have $F_\theta(x_t, y_t)-F_\tau(x_t, y_t)>0$ for any $\frac{x_t}{x_t+y_t}\in (0, \Delta_\tau) \cup (\Delta_W, 1)$, but $F_\theta(x_t, y_t)-F_\tau(x_t, y_t)<0$ for any $\frac{x_t}{x_t+y_t}\in (\Delta_\tau, \Delta_W)$ and either $F_\theta(x_t, y_t)-F_\tau(x_t, y_t)<0$ or $F_\theta(x_t, y_t)-F_\tau(x_t, y_t)>0$ when $\frac{x_t}{x_t+y_t} = \Delta_W$. Hence, $\theta$-type dominates the population as $t$ approaches infinity for any initial condition $(x_0, y_0)$ such that $\frac{x_0}{x_0+y_0}\in (\Delta_W, 1)$. When $p<\Delta_p$, $F_\theta(x_t, y_t)-F_\tau(x_t, y_t)<0$ for any $(x_t, y_t)$, implying that $\tau$-type dominates the population as $t$ approaches infinity for any initial condition $(x_0, y_0)$. When $p=\Delta_p$, $F_\theta(x_t, y_t)-F_\tau(x_t, y_t)=0$ for any $\frac{x_t}{x_t+y_t}\in (0, \Delta_\tau)\cup (\Delta_W, 1)$, implying that any state such that $\frac{x_t}{x_t+y_t} \in (0, \Delta_\tau)\cup (\Delta_W, 1)$ is a steady state. If $z_t = 1$ when $\frac{x_t}{x_t+y_t} = \Delta_W$ for all $t$ then $\frac{x_t}{x_t+y_t} = \Delta_W$ is also a steady state when $p=\Delta_p$.
%iii) Suppose $\alpha\in [\pi(C, D)+\pi(D, C)-2\pi(D, D), \frac{1}{\Delta_\tau}(\pi(C, D)+\pi(D, C)-2\pi(D, D))]$, then there exists a $\Delta_W \in [\Delta_\tau, 1)$ such that for any $\frac{x_t}{x_t+y_t} \in (\Delta_W, 1)$, $z^*_{t}=1$, and for any $\frac{x_t}{x_t+y_t} \in [\Delta_\tau, \Delta_W]$, $z^*_{t}=0$. Also, we still have $z^*_{t}=1$ for any $\frac{x_t}{x_t+y_t}<\Delta_\tau$. When $p>\Delta_p$, we must have $F_\theta(x_t, y_t)-F_\tau(x_t, y_t)>0$ for any $\frac{x_t}{x_t+y_t}\in (0, \Delta_\tau) \cup (\Delta_W, 1)$, but $F_\theta(x_t, y_t)-F_\tau(x_t, y_t)<0$ for any $\frac{x_t}{x_t+y_t}\in [\Delta_\tau, \Delta_W]$. Hence, $\theta$-type dominates the population as $t$ approaches infinity for any initial condition $(x_0, y_0)$ such that $\frac{x_0}{x_0+y_0}\in (\Delta_W, 1)$. When $p<\Delta_p$, $F_\theta(x_t, y_t)-F_\tau(x_t, y_t)<0$ for any $(x_t, y_t)$, implying that $\tau$-type dominates the population as $t$ approaches infinity for any initial condition $(x_0, y_0)$. When $p=\Delta_p$, $F_\theta(x_t, y_t)-F_\tau(x_t, y_t)=0$ for any $\frac{x_t}{x_t+y_t}\in (0, \Delta_\tau)\cup (\Delta_W, 1)$, implying that any state such that $\frac{x_t}{x_t+y_t} \in (0, \Delta_\tau)\cup (\Delta_W, 1)$ is a steady state. 
\textit{Q.E.D.}

\end{document}